\documentclass[aps,prb,twocolumn,superscriptaddress,preprintnumbers,amsmath,amssymb]{revtex4}
\usepackage{graphicx}
\begin{document}
\title{Quantum degradation of the second order phase transition}

\author{S.M. Stishov}
\email{sergei@hppi.troitsk.ru}
\affiliation{Institute for High Pressure Physics of RAS, Troitsk, Russia}
\author{A.E. Petrova}
\affiliation{Institute for High Pressure Physics of RAS, Troitsk, Russia}
\author{S.Yu. Gavrilkin}
\affiliation{P. N. Lebedev Physical Institute, Leninsky pr., 53, 119991 Moscow, Russia}
\author{L.A. Klinkova}
\affiliation{Institute of Solid State Physics of the RAS, Chernogolovka, Moscow District, Russia}

\begin{abstract}
The specific heat, magnetization and thermal expansion of single crystals of antiferromagnetic insulator EuTe, measured at temperatures down to 2 K and in magnetic fields up to 90 kOe, demonstrate non trivial properties. The Neel temperature, being $\sim9.8$ K at H=0, decreases with magnetic field and tends to zero at $\sim76$ kOe, therefore forming a quantum critical point. The heat capacity and thermal expansion coefficient reveal $\lambda$-type anomalies at the second order magnetic phase transition at low magnetic fields, evolving to simple jumps at high magnetic fields and low temperatures, well described in a fluctuation free mean-field theory. The experimental data and the corresponding analysis favor the quantum concept of effective increasing space dimensionality at low temperatures that suppresses a fluctuation divergence at a second order phase transition.
\end{abstract}
\maketitle

\section{Introduction}
As is well known, in systems undergoing second order phase transitions with well developed fluctuations, thermodynamic quantities such as heat capacity, thermal expansion coefficient, compressibility, etc. diverge or display sharp cusps at the critical point. The behavior of the thermodynamic (physical) quantities $y_1, . . . ,y_N$ in the vicinity of a phase transition point is normally described by power functions of the form $y\propto\mid\delta\mid^{-x}$, where $\delta$ is a control parameter, a dimensionless distance to the phase transition point expressed in terms of pressure, magnetic field, concentration, etc., and $x$ is the so-called critical exponent~\cite{1}.

Values of the critical exponents are defined by the universality classes, so they should not change along the phase transition line. But the question arises what would happen if the temperature of a phase transition approaches zero. In particular, it is of great interest how the diverging heat capacity and thermal expansion coefficient would agree with the Nernst theorem. This problem was studied long ago by Vaks and Geilikman, who suggested that in some cases the heat capacity at the second order phase transition may diverge down to the lowest temperature~\cite{2}. Almost simultaneously Rechester~\cite{3} and finally Khmelnitsky et al.~\cite{4} showed  that a compressibility had a weak logarithmic singularity at a second order phase transition at T=0, which implied that at temperatures just slightly higher than zero the heat capacity should be singular as well. But it should be pointed out that the theoretical analysis carried out in Ref.~\cite{2,3,4} is related to the structural phase transition of displacement type, where fluctuation effects are often suppressed or not observable even at moderate temperatures. Hence, following an evolution of these phase transitions with decreasing temperature could hardly confirm or disprove the aforementioned theories.

However, later on it was stated in a number of pioneering works~\cite{5,6,7,8} (see also~\cite{9,10}) that the effective dimensionality of quantum system in the critical region at $T=0$ might be equal or even exceed the upper critical dimension $d{_c}^{+}$ according to the relation $d_{eff}=(d+z)$, where $d$ is the space dimension and $z$ is a dynamic critical exponent. In this situation one may expect that with decreasing temperature a divergence or a sharp cusp in the heat capacity and thermal expansion coefficient in $3d$ systems would disappear and a phase transition would acquire features of the mean-field or Landau type transition. The latter is characterized by a finite jump in the thermodynamic quantities instead of their divergence at the phase transition point. Note that the above-mentioned theoretical conclusions were never properly confirmed experimentally. Though a study of variation of the order parameter exponent at low temperatures for antiferromagnetic MnCl$_2\cdot$4H$_2$0 seemingly supports the theoretical deductions on the increased effective dimensionality of quantum systems~\cite{11}.

In the present paper we report results on heat capacity, thermal expansion and magnetization measurements of the antiferromagnetic insulator EuTe in high magnetic fields.\

EuTe crystallizes in the cubic rock salt structure. Below the Neel temperature $T_N\sim9.8$ K EuTe orders antiferromagnetically with the spins lying in the (111) plane~\cite{12}. It is believed  that the  antiferromagnetic spin structure in EuTe experiences some sort of a gradual spin-flop transition in applied magnetic fields of $\sim5$ kOe, resulting in the formation of a so-called "canted" spin structure. The magnetic phase diagram of EuTe shows that on applying magnetic field the Neel temperature $T_N$ decreases and tends to zero, therefore forming the quantum critical point at $\sim76$ kOe (see~\cite{13,14}).

The heat capacity and thermal expansion coefficient of EuTe in zero magnetic field reveal sharp peaks at the phase transition point~\cite{15}.

As pointed out in Ref.~\cite{15} pure EuTe demonstrates unusual critical behavior. Considered as the 3d-Heisenberg antiferromagnet EuTe actually exhibits a positive critical exponent $\alpha$, which characterizes the behavior of the heat capacity and thermal expansion. This implies their divergence at the Neel temperature, though one should expect a negative value for the heat capacity exponent $\alpha$ (i.e., a cusp at the Neel temperature) from the 3d-Heisenberg model. However, this circumstance it is not important in the context of the present paper. Our task is to follow an evolution of the phase transition features in EuTe on approaching the quantum critical point.

\begin{figure}[htb]
\includegraphics[width=80mm]{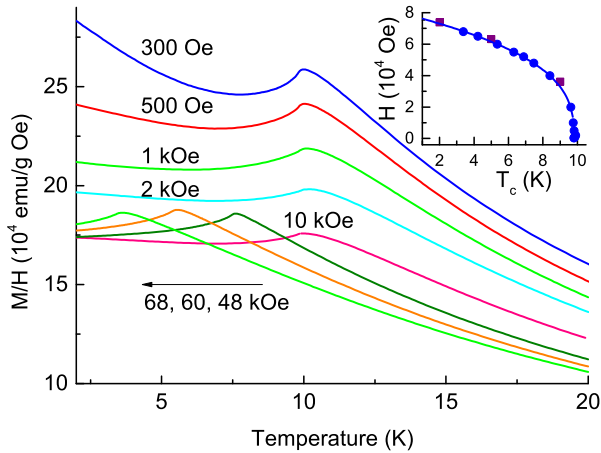}
\caption{\label{fig1} Temperature dependence of the ratio M/H at different magnetic fields for EuTe. M-magnetic moment, H-applied magnetic field. In the inset, the Curie temperature as a function of the applied magnetic field (circles taken as maxima of temperature derivatives of M/H, squares taken as points of decrease of M/H in Fig.~\ref{fig2})}
\end{figure}

\begin{figure}[htb]
\includegraphics[width=80mm]{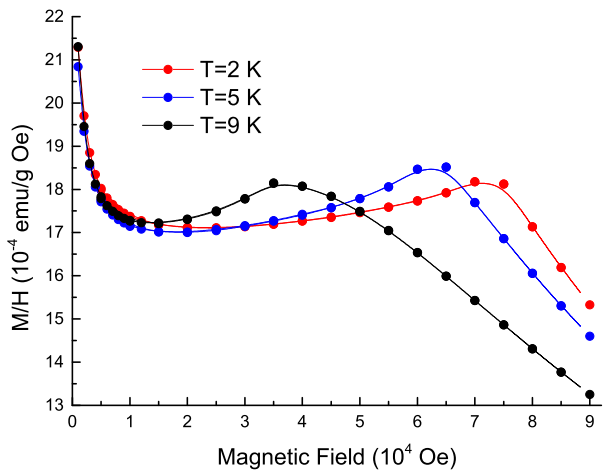}
\caption{\label{fig2} Ratio M/H as a function of applied magnetic fields for EuTe.}
 \end{figure}
\begin{figure}[htb]
\includegraphics[width=80mm]{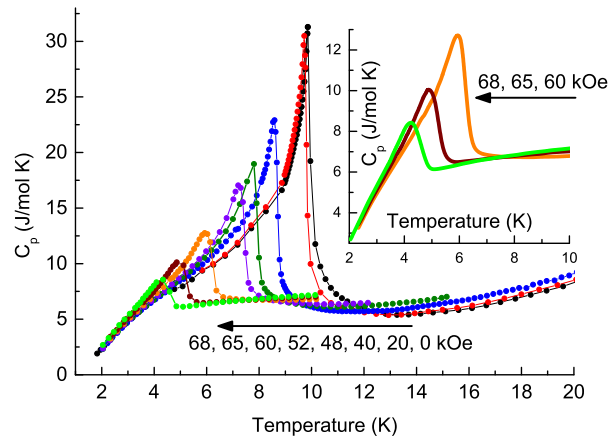}
\caption{\label{fig3} Heat capacity of EuTe across the phase transition in applied magnetic
fields.}
 \end{figure}

\section{Experimental}
Single crystals of EuTe of size about 10x10x2 mm$^3$ were grown by vapor deposition on a (100) – oriented molybdenum substrate in a molybdenum crucible~\cite{16}. The electrical resistivity of the crystals clearly exceeded a value of $10^6$ Ohm cm. The lattice parameter of the grown crystals appeared to be a=6.5993(5) \AA, which agrees well with previously published data~\cite{17}.

The specific heat, magnetization and thermal expansion were measured with a Quantum Design Physical Property Measurement System (PPMS) at temperatures down to 2 K and in magnetic fields up to 90 kOe, making use of the PPMS heat capacity module, vibration magnetometer and capacity dilatometer with a resolution of $\sim$1 \AA~\cite{18}. The dilatometer design was compatible with the PPMS system.

Results of the magnetization measurements with magnetic field  directed along [100] are depicted in Fig.~\ref{fig1}. It is seen that the ratio $M/H$ (M-magnetic moment, H-magnetic field), which is of course not a true magnetic susceptibility, strongly depends on temperature. At fixed temperature, a rapid decrease of the ratio $M/H$ at the low magnetic fields is quite evident in Fig.~\ref{fig1}.

This effect is well illustrated in Fig.~\ref{fig2}, where the selected magnetization isotherms of EuTe are displayed. The decrease of $M/H$ in the interval of 0-10 kOe  obviously reflects a gradual transition from the antiferromagnetic to a canted (spin flop) spin structure. The decrease of $M/H$ at high magnetic fields coincides with the canted-paramagnetic phase transitions (inset in Fig.~\ref{fig1}). The behavior of $M/H$ in the paramagnetic phase in Fig.~\ref{fig2} clearly exposes the saturation effects. The value of the saturation magnetization of EuTe at 2 K is equal to $\sim138$ emu/g with the magnetic field directed along [100], which is quite comparable to the data in Ref.~\cite{13}.

\begin{figure}[htb]
\includegraphics[width=80mm]{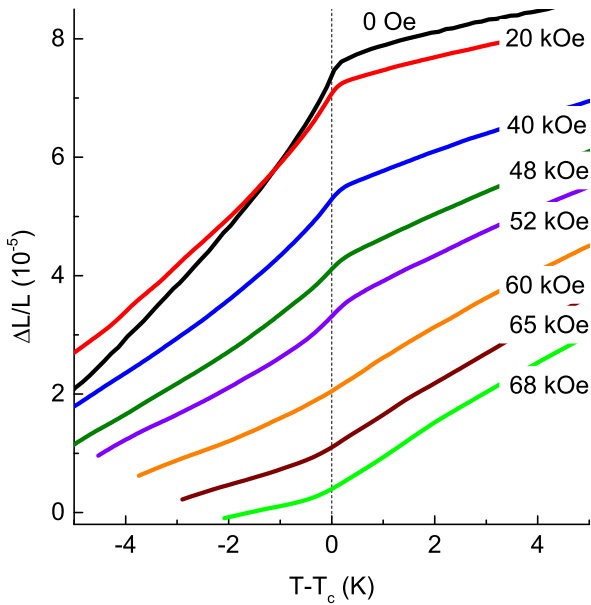}
\caption{\label{fig4} Thermal expansion of EuTe across the phase transition in applied magnetic fields.}
 \end{figure}

\begin{figure}[htb]
\includegraphics[width=80mm]{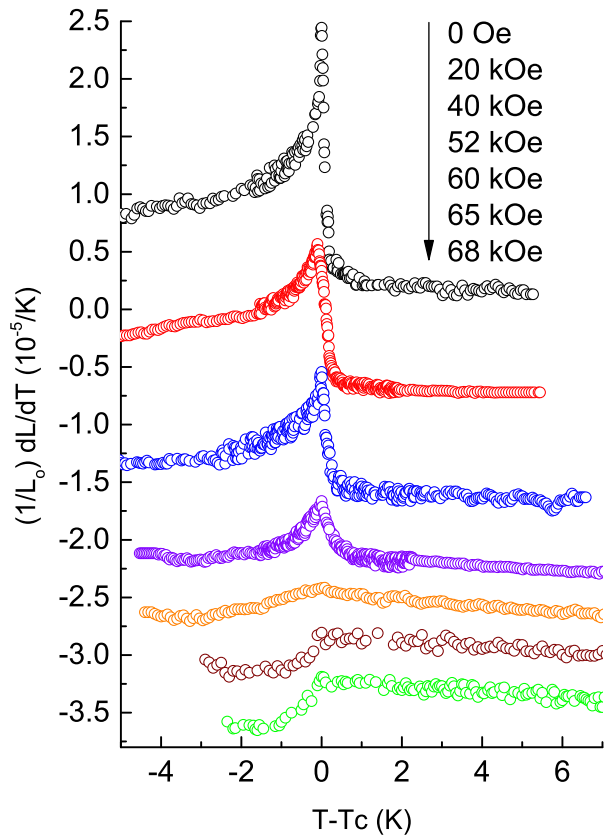}
\caption{\label{fig5} Thermal expansion coefficient of EuTe at the phase transition in applied magnetic fields. The curves for 20-68 kOe are arbitrary shifted down for a better viewing.}
 \end{figure}

Results of the heat capacity measurements are shown in Fig.~\ref{fig3}. As can be seen, the $\lambda$ - anomaly in the heat capacity, which designates the second order phase transition from antiferromagnetic (canted) to a paramagnetic spin structure, shrinks with magnetic field along the transition line, in agreements with data for Eu$_{0.95}$Sr$_{0.05}$Te~\cite{19}. That seems to be quite natural in view of gradually converting the canted spin configuration to the paramagnetic spin polarized structure on increasing the magnetic field. But new features in Fig.~\ref{fig3} are a gradual degradation of the $\lambda$-form of the heat capacity peaks, which finally take up sort of a triangular shape (see curves of 65 and 68 kOe, Fig.~\ref{fig4}). These triangle-like forms most probably originate from a discontinuity of the heat capacity as it should occur in the mean field or Landau approximation. The finite slope at the high temperature side of the mentioned curves is probably caused by sample imperfections. Moreover a trajectory of the heat capacity measurements at high magnetic fields lies almost parallel to the phase transition curve, so even a small smearing out of the transition line would enhance the effect of sample imperfections on the heat capacity. The thermal expansion of a single crystal of EuTe, as it was measured along the [100] direction, the same as the direction of the magnetic field, is shown in Fig.~\ref{fig4}. Two features are immediately observed in the figure. First, the thermal expansion curve at $H=0$, with EuTe in the antiferromagnetic phase, behaves differently from the others, measured at $H\neq0$. This obviously points to the volume effect at the gradual antiferromagnetic-canted transitions in low magnetic fields (see Fig.~\ref{fig2}). Second, the changes of slope of the thermal expansion at the phase transition strongly evolves with magnetic field and temperature, and changes sign at $\sim$60 kOe. The thermal expansion coefficient of EuTe, obtained by differentiation of the thermal expansion data, are displayed in Fig.~\ref{fig5}. It can be seen that the $\lambda$ - anomaly of the thermal expansion coefficients contracts with increased magnetic field, similar to the heat capacity anomaly. The evolution of the $\lambda$-anomaly ends with simple steps in the high-field/low-temperature limit, which clearly can be interpreted as mean field jumps slightly smeared out by sample imperfections. So the thermal expansion measurements completely support conclusions made on the basis of the heat capacity experiments.

 
\section{Discussion}
As we have seen in the previous section, the experimental data indicate remarkable change in the form of the heat capacity and thermal expansion anomalies at the antiferromagnetic phase transitions in EuTe along the transition line in the low temperature limit. At high temperatures and low magnetic field these anomalies look like fluctuation induced features typical of the majority of phase transitions. At low temperatures and high magnetic fields these anomalies reduce (degrade) to the simple step-like form associated with phase transitions with suppressed fluctuations, which normally can be described in the mean-field approximation.

So the question is what happened to the fluctuations? Where did they go? Is it a result of the quantum increase of effective dimensionality at T$_c\rightarrow 0$?  Or it is just a classical effect caused by a temperature decrease and subsequent suppression of thermal fluctuations?

To elucidate the problem let us turn to the Landau theory of phase transitions. In a simple version, the Landau theory ignores fluctuations of the order parameter and is therefore invalid in a region close to the phase transition. The simplest way to take into account fluctuations is to expand the thermodynamic potential not only in the power series of the order parameter, but also in its gradients~\cite{20}.  The first significant term of the gradient expansion is of the form
\begin{equation}\label{eq1}
\delta\Phi\propto\xi^2 \left(\frac{\partial \eta}{\partial x}\right)^2
\end{equation}
where $\eta$ is an order parameter  and the quantity $\xi$ with dimensionality of length is called the correlation length and characterizes the spatial inhomogeneity of the system.  In the Landau theory the correlation length has the Ornstein-Zernike form
\begin{equation}\label{eq2}
\xi=\xi_0(T-T_c)^{-1/2}
\end{equation}
Then, recalling relations, following from Landau theory~\cite{20} \begin{eqnarray}\label{eq3}
\eta^2=-\frac{A}{2B};\ \Delta\eta^2=\frac{T_c}{AV_c}=\frac{T_c}{A\xi^d};\nonumber\\
\Delta C_p=\frac{a^2}{2B}T_c;\ A=a(T-T_c)
\end{eqnarray}
and making use of expressions ~\ref{eq1},\ref{eq2},\ref{eq3}, one obtains for  the relative mean-square order parameter fluctuations (see also~\cite{21,22})
\begin{equation}\label{eq4}
\frac{\langle\Delta\eta^2\rangle}{\eta^2}=\frac{{T_c}^2}{\Delta C_p {\xi_0}^d(T-T_c)^{2-d/2}}
\end{equation}
where $d$ is space dimensionality.

The described approach is completely classical, but it can be used for a discussion of properties of a phase transitions at $T\rightarrow0$, because phase transitions occurring at $T>0$ can always be described in the framework of classical statistical mechanics~\cite{23, 24}. So expression~\ref{eq4} certainly captures tendencies in the behavior of a phase transition on approaching zero temperature.

It is convenient to split the right part of eq.~\ref{eq4} in two factors $a=T_c^2/\Delta C_p$ and $b=1/{\xi^d}_0(T_c-T)^{2-d/2}$. With $\Delta C_p\sim T_c$ one obtains $a={T_c}^2/{\Delta C_p}\sim T_c$. This factor would lead to a monotonous reduction of fluctuation contribution to the thermodynamics of a phase transition, which should result in decreasing the thermodynamic anomalies by a simple scaling. On the other hand, considering the factor $b$ one should take into account a variation of the bare correlation length $\xi_0$ and a change of space dimensionality $d$. For instance, an increase of the correlation length $\xi_0$ could hide fluctuation-induced features due to shortening of the temperature interval $\Delta T=T-T_c$, where fluctuations may be observed.  Phase transitions in the classical superconductors with long correlation lengths of order $10^4$ \AA\ well illustrate this situation.

In our case a change of the correlation length is highly improbable because one cannot expect a strong variation of the spin-spin interaction on the way to low temperature.

A change of space dimensionality from $d$ to $d+z$ would lead to an effective dimensionality $d_{eff}=4.5$ (dynamic exponent $z=3/2$ in the antiferromagnetic phase). In this case the relative mean-square order parameter fluctuation $\langle\Delta\eta^2\rangle/\eta^2$ will never diverge, and a phase transition will acquire mean-field features.

The change in effective dimensionality is not necessarily limited to T=0.  One should keep in mind that at $\xi_\tau < L_\tau$, where  $\xi_\tau$ - correlation time, $L_\tau =\hbar /kT$ a system is unaware of being at finite temperature and behaves if it were in a space with $d+1$ dimensions~\cite{9,10}. This implies that a transformation from $\lambda$-type anomalies to simple jumps in thermodynamic properties at a phase transition may occur at $T>0$. That is what one can see in Figs.~\ref{fig3} and \ref{fig5}.

Factor $a$, of course, influences thermodynamic anomalies as well, suppressing a fluctuation abundance. Though the latter is probably unable to change the form of an anomaly.

\section{Conclusion}
The specific heat, magnetization and thermal expansion of single crystals of antiferromagnetic insulator EuTe were measured at temperatures down to 2 K and in magnetic fields up to 90 kOe. The Neel temperature, being 9.8 K at H=0, decreases with magnetic field and tends to zero at 76 kOe, forming a quantum critical point. The heat capacity and thermal expansion coefficient reveal $\lambda$-type anomalies at the second order magnetic phase transition at low magnetic fields, which evolve to simple jumps at high magnetic fields and low temperatures, well described in a fluctuation free mean-field theory. The experimental data and the corresponding analysis favor the quantum concept of increased effective space dimensionality at low temperatures that suppress a singularity at a second order phase transition.

\section{Acknowledgements}
This work was supported by the Russian Foundation for Basic Research (grant 12-02-00376, 12-03-92604), Program of the Physics Department of RAS on Strongly Correlated Electron Systems and Program of the Presidium of RAS on Strongly Compressed Matter. We express our gratitude to Salavat Khasanov for the X-ray study of EuTe samples. The remarks and comments of Andreas Hermann are greatly appreciated.

\end{document}